\documentclass[superscriptaddress,twocolumn,showpacs,preprintnumbers,prl]{revtex4}
\usepackage{times,xspace}
\usepackage{amsbsy,amssymb,amsmath,bm}
\usepackage{graphicx,color,epsfig}
\usepackage{fancyhdr}

\def\bbbc{{\mathchoice {\setbox0=\hbox{$\displaystyle\rm C$}\hbox{\hbox
to0pt{\kern0.4\wd0\vrule height0.9\ht0\hss}\box0}}
{\setbox0=\hbox{$\textstyle\rm C$}\hbox{\hbox
to0pt{\kern0.4\wd0\vrule height0.9\ht0\hss}\box0}}
{\setbox0=\hbox{$\scriptstyle\rm C$}\hbox{\hbox
to0pt{\kern0.4\wd0\vrule height0.9\ht0\hss}\box0}}
{\setbox0=\hbox{$\scriptscriptstyle\rm C$}\hbox{\hbox
to0pt{\kern0.4\wd0\vrule height0.9\ht0\hss}\box0}}}}

\newcommand{\ket}[1]{|{#1}\rangle}
\newcommand{\bra}[1]{\langle{#1}|}

\newcommand{\ignore}[1]{}
\newcommand{\mComment}[1]{}
\newcommand{\gComment}[1]{}
\newcommand{\jComment}[1]{}
\newcommand{\rComment}[1]{}
\newcommand{\lComment}[1]{}

\newcommand{\tr}{{\rm tr}}

\renewcommand{\gComment}[1]{\textcolor{magenta}{Gerardo: #1}}
\begin{document}


\title{Condensation of Anyons  in Frustrated Quantum Magnets}




\author{C. D. Batista}
\affiliation{Theoretical Division, T-4 and CNLS, Los Alamos National Laboratory, Los Alamos, New Mexico 87545, USA.}

\author{ Rolando D.  Somma} 
\affiliation{Theoretical Division, T-4 and CNLS, Los Alamos National Laboratory, Los Alamos, New Mexico 87545, USA.}


\date{\today}

\begin{abstract}
We derive the exact ground space of a family of spin-1/2 Heisenberg chains with uniaxial exchange anisotropy (XXZ) and interactions between nearest and next-nearest-neighbor spins. The Hamiltonian family, ${\cal H}_{\rm eff}(Q)$, is parametrized by a single variable $Q$. By using a
generalized Jordan-Wigner transformation that maps spins into anyons, we show that the exact 
ground states of ${\cal H}_{\rm eff}(Q)$ correspond to a condensation of anyons with statistical 
phase $\phi=-4Q$. We also provide  matrix-product state representations of  some ground states 
that allow for the efficient computation of spin-spin correlation functions.
\end{abstract}

\pacs{%
}

\maketitle


One-dimensional quantum magnets  can realize exotic states of matter such as Luttinger liquids~\cite{1.,2.}, valence bond solids~\cite{3.,4.}, 
and spin supersolids~\cite{5.}. A unique feature of these systems is that transmutations of particle statistics  preserve the range  of the interactions. This is the main reason
behind the success of spin-fermion transformations for studying
 one-dimensional spin models~\cite{6.,7.,8.}. A simple generalization
of the Jordan-Wigner transformation~\cite{6.} allows us to map spins into anyons (particles
that generalize the concept of fermions and bosons). By exploiting this generalization,  we  
derive exact ground states of  spin-1/2 
frustrated magnets and provide an efficient method for computing spin-spin correlation functions. Remarkably, the resulting ground  states  are {\em anyon condensates}
that spontaneously break the particle-number conservation symmetry. 
In contrast to the familiar Bose-Einstein condensates (BECs), the condensed particles satisfy  anyonic statistics. 

BECs are ubiquitous in many-body physics. Atomic gases, certain quantum magnets, and superconductors can be  
described as interacting gases of bosons that become BECs at low enough temperatures~\cite{9.,10.,11.,12.,13.}. 
A BEC  is characterized by the spontaneous breaking of the symmetry associated 
with the boson-number conservation. This  symmetry breaking comes along with a  macroscopic occupation of the  single-particle ground state, suggesting that the notion of  BEC  cannot be generalized to other particle statistics. Indeed, {\it anyons}, which are identical particles obeying arbitrary statistics ~\cite{14.}, are subject to a generalized Pauli exclusion principle that limits the occupancy of any given state. Therefore, the only way of generalizing the notion of BECs to  {\it anyon condensates} (ACs) is by taking an adequate limit over the particle statistics that survives the exclusion principle. Like BECs, ACs spontaneously break  the particle-number conservation symmetry.

Just as BECs can only exist at zero-temperature in  low-dimensional systems with short range interactions~\cite{15.}, we will demonstrate that ACs are the  ground states
of a family of one-dimensional frustrated quantum magnets.  
The anyonic statistics is fixed by a phase 
$\phi$ that their wave function acquires after  exchanging  two anyons: $\Psi({\bf r}_1, {\bf r}_2) = e^{i\phi} \Psi({\bf r}_2, {\bf r}_1)$. This condition implies that anyons are in general ``non-local'' particles that  interpolate continuously between bosons ($\phi=0$)  and fermions ($\phi=\pi$).
 
\begin{figure}[ht]
  \centering
  \includegraphics[width=7.2cm]{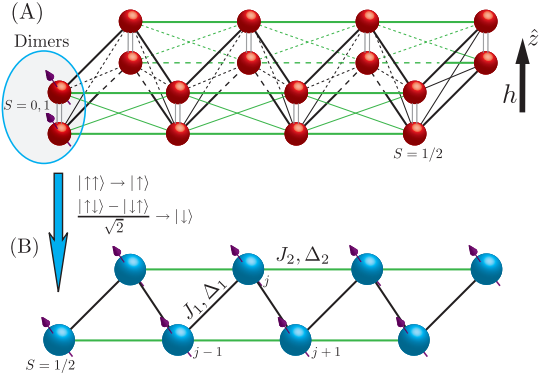}
  \caption{Frustrated quantum magnet. (A) Original spin-1/2 system: 
pairs of spin-1/2 particles in the same dimer interact via dominant antiferromagnetic
exchange (double lines). Weaker inter-dimer exchange interactions are represented by single lines. 
(B) Effective spin-1/2 XXZ model, ${\cal H_{\rm eff}}$, that results from projecting the original Hamiltonian into the low-energy subspace.}
  \label{fig:effectivemodel}
\end{figure}

We will start by considering  the dimerized $S=1/2$ spin
system  of Fig.~\ref{fig:effectivemodel} A. Pairs of spin-1/2 particles in the same dimer interact via dominant antiferromagnetic exchange (double lines). Weaker inter-dimer exchange interactions are represented by single lines. In presence of an external field $h$, the low-energy single-dimer states are the triplet   $\ket{\uparrow \uparrow}$ ($s^z=1/2$) and the singlet $\ket{\uparrow \downarrow} - \ket{\downarrow \uparrow}$  ($s^z=-1/2$).  
The low-energy effective spin-1/2 XXZ model, ${\cal H_{\rm eff}}$, that results from projecting the original Hamiltonian into the low-energy subspace is ~\cite{16.,17.,18.,19.} (see Fig.~\ref{fig:effectivemodel}):
\begin{eqnarray}
\nonumber
{\cal H}_{\rm eff} =\!\!\!\sum_{j;\nu=1,2} J_{\nu} \left [ \Delta_\nu \left ( s^z_{j+\nu} s^z_j - \frac{1}{4} \right)
+ s^x_{j+\nu} s^x_{j} + s^y_{j+\nu} s^y_{j}  \right].
\label{eq:Ham}
\end{eqnarray}
The eigenstates of $s^z_j$ represent the singlet  and the triplet of the $j$-th dimer Hamiltonian. 

In this Letter we  derive the exact ground states of  ${\cal H}_{\rm eff}(Q)$ for  
$\Delta_\nu = \cos (\nu Q)$ and $J_1 = -4J_2 \cos Q$, with $0 \le Q <\pi$.
If $Q= \pi/2$ the model reduces to two decoupled ferromagnetic Heisenberg chains whose
exact solutions are known~\cite{20.}. Consequently, if the number of effective spins in the ladder is $L$,
the ground-space dimension is $(L/2+1)^2$ or $(L+3)(L+1)/4$ for even or odd $L$. 
We will prove that the ground-space dimension is independent
of  $Q$ for a special choice of open boundary conditions.

\begin{figure}[ht]
  \centering
  \includegraphics[width=7.1cm]{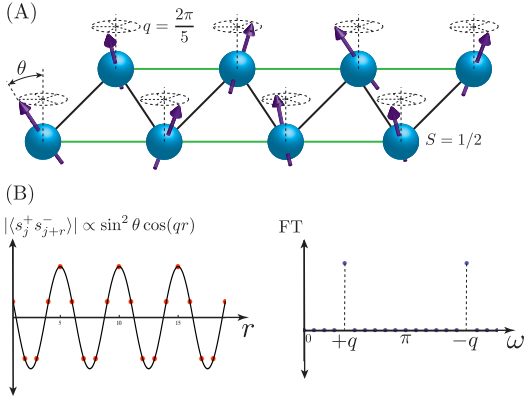}
\caption{Canted spiral order.
(A) Spin-1/2 representation of the spiral states $\ket{s_\theta (Q)}$ for $Q=2\pi/5$. 
(B) Spin-spin correlation as a function of distance $r$.
}
 \label{fig:spiralorder}
\end{figure}

Some of the ground states are product states $\ket{s_{\theta} (Q)}$, in which the spins are aligned according to a canted spiral
of wave vector  $Q$ and arbitrary canting angle $\theta$~\cite{21.} (see Fig.~\ref{fig:spiralorder} A).
$\ket{s_{\theta} (-Q)}$ are also ground states due to spatial inversion symmetry: ${\cal H}_{\rm eff} (Q)={\cal H}_{\rm eff} (-Q)$.
The long-range spiral order is manifested by the spatial dependence of spin-spin correlations that oscillate with constant amplitude and wave-vector $\pm Q$  (see Fig.~\ref{fig:spiralorder} B). 

States with spiral order can be represented as BECs
by mapping 
spin-1/2 operators into hard-core bosons  via the Matsubara-Matsuda (MM) transformation~\cite{22.}:
$b^{\dagger}_j =s^+_j$. The operator
$b^{\dagger}_j $ creates a hard-core boson at the $j$-th site of the ladder.
The identity $(b^\dagger_j)^2=0$ excludes  states with more
than one boson per site.
Since ${\cal H}_{\rm eff} (Q)$ is translationally invariant,  
it is useful to introduce the operators
$\bar b^{\dagger}_k = \sum_{j=1}^L \exp (i k j ) b^{\dagger}_j /\sqrt L$
that create bosons of momentum $k$. 
Contrary to  real space bosons, $b^\dagger_j$, occupancy in momentum space is not subject to exclusion.
The BEC states
\begin{equation}
\label{eq:BECs}
\ket{\psi_n (\pm Q)} \propto (\bar b^\dagger_{\pm Q})^n \ket{\varnothing} \; 
\end{equation}
span the ground space with spiral order~\cite{21.}.
$\ket{\varnothing} $ is the vacuum or empty state
that corresponds to  the fully-polarized state $\ket{\downarrow \ldots \downarrow}$ in the  spin language
and $n$ is the number of single bosons with momentum $\pm Q$.
 $\ket{\psi_n ( Q)}$ and $\ket{\psi_n (- Q)}$ are usually non-orthogonal but linearly independent. Finding all BEC states for
$Q=0$ requires to consider adequate linear combinations of  $\ket{\psi_n ( Q)}$ and $\ket{\psi_n (- Q)}$ before taking the limit $Q \rightarrow 0$. The single-boson excited eigenstates, $\bar b^\dagger_k \ket{\varnothing}$,   have energy $\omega_k = 2 J_2 (\cos k - \cos Q)^2$~\cite{21.}.

The MM transformation provides then a  simple alternative representation
of the spiral ground states in terms of hard-core bosons.
Remarkably, 
a generalization of the particle statistics will allow us to find
exact  expressions for {\em all}  ground states of ${\cal H}_{\rm eff} (Q)$.
The generalized JW transformation
$a^\dagger_j = \exp\{i \phi \sum_{l<j} (s^z_l+1/2)\} s^+_j$
maps spins into anyons, 
where the operator $a^\dagger_j $  creates a hard-core anyon [$(a^\dagger_j)^2=0$] at the $j$-th site of the ladder.
The MM and JW transformations are recovered for $\phi =0$ (hard-core bosons)
and  $\phi=\pi$ (spinless fermions), respectively.
The anyonic statistics becomes manifest in the commutation relations $a^{\dagger}_j a^{\dagger}_l = e^{-i \phi} a^{\dagger}_l a^{\dagger}_j$, when $j <l$. 
The states
\begin{equation}
\label{eq:ACs}
\ket{\psi_{n,m}( Q)} = \lim_{\phi \rightarrow  -4Q} c_{ \phi}
(\bar a^\dagger_Q)^n (\bar a^\dagger_{-Q})^{m} \ket{\varnothing} \; , 
\end{equation}
with $0 \le n,m \le L/2$, provide a natural generalization of the BEC states in Eq.~\eqref{eq:BECs}.
  $\bar a^{\dagger}_k = \sum_{j=1}^L \exp (i k j ) a^{\dagger}_j/\sqrt L$ is the operator
that creates an anyon with momentum $k$
and  $c_\phi$ is the normalization constant.  The limit in Eq.~\eqref{eq:ACs} is necessary to avoid 
a trivial cancellation due to a generalized Pauli exclusion principle: $(\bar a^{\dagger}_k)^p=0$ for $\phi= 2\pi /p$ and $p$ an integer greater than 1.
The statistical phase does not play a  role if  $n=0$ or $m=0$ because $\lim_{\phi \rightarrow -4Q} c_\phi (\bar a^\dagger_Q)^n \ket{\varnothing} \propto (\bar b^\dagger_Q)^n \ket{\varnothing}$. However, the same limiting procedure leads to states
that are qualitatively different from a double-$Q$ BEC, $(\bar b^\dagger_Q)^n(b^\dagger_{-Q})^m \ket{\varnothing}$, when $n,m \neq 0$.

\begin{figure}[ht]
  \centering
  \includegraphics[width=7.2cm]{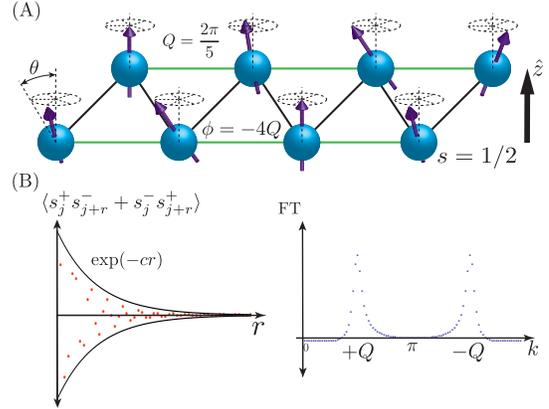}
  \caption{Phase soliton. (A)
A topological defect results from the creation of an anyon  
on the spiral  state, i.e.  $a^\dagger_j \ket{s_\theta(-Q)}$ (Fig.~\ref{fig:spiralorder} A).
(B) Exponential decay of  spin-spin correlations  as a function of distance $r$ 
for an AC. This correlation function was obtained by using a matrix-product state representation.
The Fourier transform (FT) of the correlation function is peaked at the momentum of the AC.
$c>0$, $Q=2\pi/5$, and $L=100$.
}
  \label{fig:disorder}
\end{figure}

$\ket{\psi_{n,m}( Q)}$ span the full ground space of ${\cal H}_{\rm eff} (Q)$ -- see below. 
As $n+m$ is not preserved, the particle-number symmetry is spontaneously broken.
Therefore, we  refer to the physical phenomenon characterized by such states
as {\em anyonic condensation} and to  $\ket{\psi_{n,m}( Q)}$  as {\em AC states}.
The physical properties of ACs are better understood if we first consider  the action
of  $a^\dagger_j$. Besides polarizing the $j$-th spin,
such operator creates a ``phase kink'':  it rotates the spins on the {\em left} of $j$ by an angle $\phi$ along the $\hat z$ axis. The phase kink gives rise to the anyonic statistics  that is absent in the MM transformation.
Consequently, the ground state $\bar a^\dagger_Q \ket{s_\theta (-Q)}$ 
corresponds to a phase soliton attached to a boson that propagate with momentum $Q$
across the spiral -- see Fig.~\ref{fig:disorder} A. In general, the ground states $(\bar a^\dagger_Q)^n \ket{s_\theta (-Q)}$
will correspond to a condensation of solitons.

To demonstrate that $\ket{\psi_{n,m}( Q)}$ is a ground state of ${\cal H}_{\rm eff} (Q)$, we rewrite the  ${\cal H}_{\rm eff} (Q)$ as
\begin{align}
\label{eq:Hprojector}
{\cal H}_{\rm eff} (Q) = \sum_{j=2}^{L-1} (J_2+J_1^2/(8J_2)) \Pi_j(Q) \; ,
\end{align}
where the $\Pi_j (Q)$ are projectors acting on triangular plaquettes~\cite{21.}.
These projectors are
\begin{align}
\nonumber
\Pi_j (Q)= \ket{\xi_j^\uparrow} \bra{\xi_j^\uparrow} + \ket{\xi_j^\downarrow} \bra{\xi_j^\downarrow} \; ,
\end{align}
where $\ket{\xi_j^\uparrow} \propto (2 \cos Q s^+_{j-1} s^+_{j+1} - s^+_j s^+_{j+1} - s^+_{j-1} s^+_j) \ket{\downarrow \downarrow \downarrow}$ and $\ket{\xi_j^\downarrow}  \propto  (2 \cos Q s^+_{j}  -   s^+_{j-1} - s^+_{j+1} ) \ket{\downarrow \downarrow \downarrow}$ are plaquette states.
Note that Eq.~\eqref{eq:Hprojector} refers to a particular choice of open boundary conditions. 
Since the fully-polarized spin states $\ket{\downarrow \ldots \downarrow}$ and $\ket{\uparrow \ldots \uparrow}$ are orthogonal to $\ket{\xi_j^\sigma}$, they are zero-energy ground states of ${\cal H}_{\rm eff} (Q)$.
This implies
that ${\cal H}_{\rm eff} (Q)$ is a so-called parent Hamiltonian and any other 
 ground state must have zero energy.

The AC states satisfy
\begin{align}
\nonumber
\ket{\psi_{n,m}(Q)}  =  \lim_{\phi \rightarrow -4Q} c_{\phi} (\bar a^\dagger_Q)^n (\bar a^\dagger_{-Q})^m \ket{\varnothing} \\
\label{eq:anyonicdec}
 \propto \sum_{\begin{matrix}
j_1<\ldots <j_n \cr l_1<\ldots <l_m
\end{matrix}
} 
\prod_{q=1}^n e^{iQj_q} a^\dagger_{j_q} \prod_{r=1}^m e^{-iQ l_r} a^\dagger_{l_r} 
 \ket{\varnothing} \; .
\end{align}
The limit in Eq.~\eqref{eq:anyonicdec}   removes the prefactors that arise from  exchanging 
 two anyons with same momentum. 
 This is necessary to avoid
a trivial cancellation of $(\bar a^\dagger_{\pm Q})^n$ caused by the generalized
exclusion principle. 
In particular, the limit implies that 
 $\ket{\psi_{n,0} (Q)} \propto (\bar b^\dagger_Q)^n \ket{\varnothing}$ and
  $\ket{\psi_{0,m} (Q)} \propto (\bar b^\dagger_{-Q})^m \ket{\varnothing}$.
Let  $v_p= \{ p-1,p,p+1\}$. 
Then, if  $j_q \notin v_p$ and $l_r \notin v_p$ for all $q,r$,
\begin{equation}
\nonumber
\Pi_p(Q) \prod_{q=1}^n   a^\dagger_{j_q} \prod_{r=1}^m  a^\dagger_{l_r}  \ket{\varnothing}=
\prod_{q=1}^n   a^\dagger_{j_q} \prod_{r=1}^m   a^\dagger_{l_r}  \Pi_p(Q)  \ket{\varnothing}=0.
\end{equation}
We now consider the terms in Eq.~\eqref{eq:anyonicdec}  for which a single $j_i \in v_p$ and  $l_r \notin v_p$
for all $r$.
We fix all the indexes except for $j_i$  and consider the sum over the terms 
that have $j_i$ in $v_p$. The resulting state satisfies
\begin{align}
\nonumber
& \Pi_p(Q) \sum_{j_i=p-1}^{p+1}  \prod_{q = 1}^n e^{iQj_q} a^\dagger_{j_q} \prod_{r=1}^m e^{-iQ l_r} a^\dagger_{l_r}  \ket{\varnothing}  \propto 
\\
\nonumber 
& \prod_{q \ne i } e^{iQj_q} a^\dagger_{j_q} \prod_{r=1}^m e^{-iQ l_r} a^\dagger_{l_r} \Pi_p(Q)   \sum_{j_i=p-1}^{p+1} e^{iQ j_i} a^\dagger_{j_i} \ket{\varnothing}  =0 \; .
\end{align}
This equality follows from $ \sum_{j_i=p-1}^{p+1} e^{iQ j_i} a^\dagger_{j_i} \ket{\varnothing} $
being orthogonal to $\ket{\xi_p^\sigma}$.
A similar reasoning applies if a single $l_r$ is in $v_p$ and $j_q \notin v_p$ for all $q$.
Finally, we analyze the case for which
only two indexes of $\{ j_q \}_q \cup \{ l_r \}_r$ are in $v_p$. If the two indexes are either in $\{j_q \}_q$
or $\{ l_r \}_r$, we can use a similar argument as the one above to show that $\Pi_p(Q)$ acts trivially in the particular sum of states. We  then analyze the case
for which one of the indexes in $v_p$ is in $\{j_q\}_q$ and the other one
 is in
$\{l_r\}_r$. We label these indexes by $j_i$ and $l_{h}$, respectively. 
If we fix all the remaining
indexes and  consider the sum of terms over the six possible values of $(j_i,l_h)$, the resulting
state satisfies
\begin{align}
\nonumber
& \Pi_p(Q) \sum_{(j_i,l_h)}  \prod_{q = 1}^n e^{iQj_q} a^\dagger_{j_q} \prod_{r=1}^m e^{-iQ l_r} a^\dagger_{l_r}  \ket{\varnothing} \propto 
\\
\nonumber
&  \prod_{q  \ne i} e^{iQj_q} a^\dagger_{j_q} \prod_{r \ne h} e^{-iQ l_r} a^\dagger_{l_r} 
H_p(Q) \sum_{(j_i,l_h)} e^{iQ(j_i-l_h)} a^\dagger_{j_i} a^\dagger_{l_h} \ket{\varnothing} \; .
\end{align}
The state $\sum_{(j_i,l_h)} e^{iQ(j_i-l_h)} a^\dagger_{j_i} a^\dagger_{l_h} \ket{\varnothing}$
is orthogonal to $\ket{\xi_p^\sigma}$ when
\begin{align}
\nonumber
 e^{iQ} +  e^{i(3Q+\phi)}  
-  \cos Q (1 + e^{i(4Q + \phi)})  = 0 \; ,
\end{align}
where $\phi$ is the anyonic statistical phase. This condition is satisfied for $\phi=-4Q$.

For even $L$, ${\cal H}_{\rm eff}(\pi/2)$ corresponds to two decoupled ferromagnetic Heisenberg chains of length $L/2$. The ground space is spanned by the linearly independent states
$\ket{\psi_{n,m}(\pi/2)}$, with $0 \le n,m \le L/2$,
and the ground space dimension is $(L/2+1)^2$. Because the AC states are analytic in $Q$,
the set $\{ \ket{\psi_{n,m}( Q} \}_{n,m}$ remains linearly independent and the dimension 
of the corresponding subspace does not change with $Q$. (Numerical calculations
suggest that these are all possible ground states.) 
The states $\ket{\psi_{n,m}( Q)}$ and $\ket{\psi_{m,n}(-Q)}$ become equal for $Q\rightarrow 0$. 
However, it is possible to obtain two linearly independent states by
taking a proper limit of linear combinations of $\ket{\psi_{n,m}( Q}$  and $\ket{\psi_{m,n}( Q}$.
For instance,
\begin{align}
\nonumber
& \lim_{Q \rightarrow 0} \frac 1 Q \left[ \ket{\psi_{n,0}(Q) }-\ket{\psi_{0,n}(Q) }\right]
\propto 
\\
\nonumber
& \propto \sum_{j_1<\ldots <j_n}(j_1+\cdots +j_n) b^\dagger_{j_1}\ldots b^\dagger_{j_n} \ket{\downarrow \ldots \downarrow}
\end{align}
 is also  a ground state of ${\cal H}_{\rm eff}(0)$.

The difference between ACs and BECs is evident from their single-particle correlation functions. 
The two-point correlator exhibits long-range order,
$\lim_{r \rightarrow \infty }|\langle s^{+}_j s^{-}_{j+r}\rangle |=\lim_{r \rightarrow \infty }| \langle b^{\dagger}_j b^{\;}_{j+r} \rangle| \ne 0$, for a BEC
(Fig.~\ref{fig:effectivemodel} B). 
In the anyonic language,
the operator $s^+_j s^-_{j+r}$ maps to $ \exp\{i \phi N_{jr}\}
a^\dagger_j a^{\;}_{j+r}$, where
$N_{jr} = \sum_{j \le l < j+r} a^\dagger_l a^{\;}_l$
  counts the number of anyons between sites $j$ and $j+r$.
  Then,
  the large fluctuations of $N_{jr}$ present in an AC translate into large fluctuations
  of the relative azimutal phase between spins $j$ and $j+r$ -- see Fig.~\ref{fig:disorder} A. 
If we assume that the probabilities of having anyons in different sites are roughly independent of each other,
we obtain $\langle s^+_j s^-_{j+r} \rangle \sim e^{-iQr}(1-\rho (1- e^{i\phi}) )^r $ for large $r$ and $\rho=n/L$. 
This simple analysis suggests that the two-point correlator in the ACs
  decays exponentially in $r$ -- see Fig.~\ref{fig:disorder} B.

Spin-spin correlations can be efficiently computed by introducing matrix-product state representations for the ground states that are invariant under translation~\cite{23.,24.}.
These states are
represented as
\begin{eqnarray}
\label{eq:MPSrep}
\ket{\psi (Q)} & \propto & \! \! \!  \sum_{\sigma_1, \ldots ,\sigma_L }
\tr \left [ A^{\sigma_1}(Q) \ldots A^{\sigma_L}(Q) \right] \ket{\sigma_1 \ldots \sigma_L},
\end{eqnarray}
where each $A^{\sigma_j}(Q)$ is a matrix associated with the $j$-th spin in the ladder
and $\sigma_j \in \{\uparrow, \downarrow \}$.
Correlation functions can be computed analytically or numerically in this representation by
a contraction of a tensor network~\cite{25.}.
The translational-invariant ground states of the matrix-product form of
 Eq.~\eqref{eq:MPSrep} satisfy
\begin{align}
\label{eq:MPSproperty}
\Pi_p(Q) \ket{\psi(Q)} =0 \ \forall \ p \; .
\end{align}
The action of $\Pi_p(Q)$ on $\ket{\psi(Q)}$
outputs another matrix-product state
whose three-spin tensor, associated with $v_p$, is
\begin{align}
\label{eq:3spintensor}
& B^{\tau_{p-1},\tau_p,\tau_{p+1}} (Q)  = 
\\
\nonumber
&= \sum_{\sigma_{p-1},\sigma_p,\sigma_{p+1}} 
A^{\sigma_{p-1}}  (Q) A^{\sigma_p}  (Q) A^{\sigma_{p+1}}  (Q) \times
\\
\nonumber
& \times \bra{\tau_{p-1},\tau_p,\tau_{p+1}} \Pi_p(Q) \ket{\sigma_{p-1},\sigma_p,\sigma_{p+1}} , 
\end{align}
and $\tau,\sigma \in \{\uparrow, \downarrow \}$. Equation~\eqref{eq:3spintensor} follows from the contraction of a tensor network~\cite{25.}. $B ^{\tau_{p-1},\tau_p,\tau_{p+1}} (Q)=0$
for all $\tau_{p-1},\tau_{p},\tau_{p+1}$ is 
a sufficient condition to satisfy Eq.~\eqref{eq:MPSproperty}.
It follows from the definition of $\ket{\xi^\sigma_p}$ that:
\begin{align}
\nonumber
A^\uparrow(Q) & A^\downarrow (Q) A^\downarrow(Q) + A^\downarrow (Q) A^\downarrow (Q) A^\uparrow (Q) 
-
\\
\nonumber
& - 2 \cos (Q) A^\downarrow (Q) A^\uparrow(Q) A^\downarrow (Q)
 = 0 \; , \\
 \nonumber
A^\downarrow(Q)  & A^\uparrow(Q)  A^\uparrow(Q)  + A^\uparrow(Q) A^\uparrow (Q) A^\downarrow(Q) -
 \\
\label{eq:MPS}
& - 2 \cos( Q) A^\uparrow(Q) A^\downarrow (Q)  A^\uparrow(Q)
 = 0 \; .
\end{align}
These equations are satisfied if $A^\uparrow (Q) A^\downarrow (Q) = e^{\pm iQ }A^\downarrow (Q) A^\uparrow(Q) $. Equations~\eqref{eq:MPS} are also satisfied for $Q=2\pi/3$  if $A^{\uparrow,\downarrow}(2\pi/3)$ are the matrices that correspond to a ground state of the well-known Majumdar-Ghosh model~\cite{26.}.

Solutions for $Q = 2 \pi/p$, with integer $p$, are given by
$A^\uparrow(Q)=Z_p$ and $A^\downarrow(Q) = (aX^{\;}_p + bX^\dagger_p) Z^\dagger_p$.
The complex constants $a$ and $b$ are arbitrary and
the corresponding states span the full translational-invariant ground subspace.
$Z_p$ and $X_p$ are $p$-dimensional unitary
matrices that 
satisfy $Z_p X_p = e^{-iQ} X_p Z_p$. This condition resembles the anyonic nature
of our solutions whose statistical phase is $\phi=-4Q$.
 For example, $Z_p$ is the diagonal matrix and $X_p$
is the cyclic permutation that
generate the finite Heisenberg or generalized Pauli group of order $p^2$~\cite{27.}:
\begin{align}
\nonumber
 Z_p = \begin{pmatrix} 1 & 0 & 0 & \cdots & 0 \cr 0 & \omega & 0 & \cdots & 0  \cr
 \vdots  & \vdots & \ddots & \cdots & \vdots \cr 0 & 0 & 0 & \cdots & \omega^{p-1} \end{pmatrix}  ,
 \ X_p = \begin{pmatrix} 0 & 1 & 0 & \cdots & 0 \cr 0 & 0 & 1 & \cdots & 0 \cr
 \vdots & \vdots & \vdots & \ddots & \vdots \cr 1 & 0 & 0 & \vdots & 0 \end{pmatrix}  ,
 \end{align}
 and $\omega= \exp (i2\pi/p)$.
Few-spin correlators can be numerically obtained with
a low computational cost that is polynomial in $p$~\cite{24.}. 
For instance,
\begin{align}
\nonumber
\bra{\psi(Q)}  s^+_j & s^-_{j+r}  \ket{\psi(Q)} =   \tr \left[ E^{j-1} \times (A^\downarrow(Q) \otimes  (A^\uparrow(Q))^*) \times \right. \\
\label{eq:spin-spincorrel}
& \left. \times E^{r-1} \times
(A^\uparrow(Q) \otimes  (A^\downarrow(Q))^*) \times E^{L-r} \right] \; ,
\end{align}
and $E= A^\uparrow(Q) \otimes (A^\uparrow(Q))^* + A^\downarrow(Q) \otimes (A^\downarrow(Q))^*$.
The normalization constant is $\bra{\psi(Q)}  \psi(Q) \rangle = \tr [E^L]$.
We used Eq.~\eqref{eq:spin-spincorrel} and Gram-Schmidt orthogonalization  to 
compute the exact spin-spin correlator in Fig.~\ref{fig:disorder} B and  confirm the
expected exponential decay for ACs.

Finally, it is interesting to note that the particular ground state solutions \eqref{eq:BECs} imply that the scattering length of two bosons with momenta
$k$ and $k'$ goes asymptotically to zero for $k, k' \to Q$ or $k,k' \to -Q$. However, the boson-boson scattering length remains
finite if $k \to Q$ and $k' \to {-Q}$. In contrast, our general ground state solutions given in Eq.~\eqref{eq:ACs}  imply that the scattering length of two {\it anyons} with momenta $k$ and $k'$ goes asymptotically to zero
in both cases: $(k,k') \to (Q,Q)$ or  $(k,k') \to (- Q, -Q)$  {\it and}  $(k,k') \to (Q, -Q)$. This implies that  the
low-energy excitations on top of the condensate  are quasi-free anyons.

In summary, we have introduced the notion of anyon condensates as a generalization of BECs.
By using a natural extension of the  Jordan-Wigner transformation, 
we have shown that that the AC states \eqref{eq:ACs} are  the exact ground states of a family 
one-dimensional XXZ magnets with nearest and next-nearest-neighbor exchange interactions.
 ACs  spontaneously break the particle-number conservation symmetry and correspond to a condensation
of topological defects.
We  have also provided a  matrix-product representation for the translational-invariant ground states, 
which allows for the efficient computation of  correlation functions. 
These  states also satisfy an {\em area law}, which is rather common in frustrated systems, 
and results in an upper bound on the entanglement entropy~\cite{28.}.
The exponential decay obtained for the single-particle correlator in the AC states is a natural consequence of
the topological nature of anyons. Since each anyon is a bound state of a boson and a phase soliton, the large 
fluctuations in the anyon number that characterize an AC state lead to large phase fluctuations. This behavior is 
qualitatively different from the bosonic limit (BEC) in which particle number and phase are conjugate variables and
large number fluctuations lead to long-range phase coherence.

\begin{acknowledgments}
Work at LANL was performed under the auspices of the U.S.\ DOE contract No.~DE-AC52-06NA25396 through the LDRD program.
\end{acknowledgments}


\end{document}